# Entanglement of the Two Movable Mirrors in a Single Mode Cavity Field

Yong-Yi Huang

*MOE Key Laboratory for Nonequilibrum Synthesis and Modulation of Condensed Matter,
Department of Optic Information Science and Technology,
Xi'an Jiaotong University, Xi'an 710049, China*

**Abstract**

The entanglement of the two movable mirrors in a single mode cavity field is studied in an ideal situation without thermal noise. When $gn/\omega \ll 1$ where $g$ is the coupling coefficient between the mirror and the field, $\omega$ is the mirror' vibration frequency and $n$ is photon number of the cavity field, we work out Wootters concurrence of the system initially with the field in Fock number state and the two mirrors in vacuum states, and find that the states of the mirrors are entangled into the space spanned by $|0>, |1>$ via radiation pressure, though the concurrence is very small.

**Key words**: entanglement of two mirrors, single mode cavity field, vacuum states

**PACS**: 03.67.Bg, 03.65.Ud, 42.50.Pq

## 1. Introduction

The scheme of laser interferometer gravitational-wave observatory[1] attracts people much interest to study optomechanical coupling via radiation pressure, for example[2]. Using dynamical backaction a mechanical oscillator is in theory even cooled to ground state to enhance the sensitivity of displacement measurements[3,4]. Quantum entanglement is a fundamental phenomenon[5,6], it not only provides insight into the fundamental structure[7] but also become a basic resource for many quantum information processing schemes[8]. So entanglement of optomechanical coupling should be paid attention to. Actually the entanglement between a movable mirror and a cavity field is proposed[9] driven by a laser. It is surprising that macroscopic movable mirrors in two cavities are entangled via radiation pressure after considering the effect of thermal noise with driving of an intense classical field [10,11] or without driving of a classical field[12,13,14]. What's more, the entanglement of the two movable mirrors respectively in two weak coherent cavity fields is generated after the detection of a photon leaking from the cavities[15]. In this paper we show that the entanglement of the two mirrors in only one cavity is also generated even without considering thermal noise and classical field driving. The condition of entanglement is $gn/\omega_m \ll 1$, where $g$ is the coupling coefficient between the mirror and the field, $\omega_m$ is the mirror' vibration frequency and $n$ is photon number of the cavity field.

## 2. Theory Model

We study the interaction between two movable mirrors and one cavity field. Regarding two

movable mirrors as quantum harmonic oscillators, we write the system Hamiltonian as[16]

$$H = \hbar\omega b_1^+ b_1 + \hbar\omega b_2^+ b_2 + \hbar\omega_0 a^+ a - \hbar g a^+ a(b_1^+ + b_1) - \hbar g a^+ a(b_2^+ + b_2). \quad (1)$$

Here $\omega_0$ is the cavity field mode frequency, $\omega$ is the movable mirror frequency, $g$ is optomechanical coupling coefficient and $g = \frac{\omega_0}{L}\sqrt{\frac{\hbar}{2m\omega}}$ with L being the length of the cavity, $m$ being mass of the movable mirror. $a^+, b_i^+$ in equ. (1) are respectively creation operator for the cavity mode and the movable mirrors. The time evolution operator of the system is given by similarly following [17]

$$U(t) = e^{-itra^+a} e^{2i(ka^+a)^2(t-\sin t)} e^{ka^+a(\eta b_1^+ - \eta^* b_1 + \eta b_2^+ - \eta^* b_2)} e^{-it(b_1^+ b_1 + b_2^+ b_2)}, \quad (2)$$

where $r = \omega_0/\omega$, $k = g/\omega$ and $\eta = 1 - e^{-it}$, t is a scaled time, being the actual time multiplied by $\omega$, i.e. $t \equiv \omega \times t_{actual}$. The time evolution operator in the interaction picture relative to the free cavity field is

$$U_I(t) = e^{2i(ka^+a)^2(t-\sin t)} e^{ka^+a(\eta b_1^+ - \eta^* b_1 + \eta b_2^+ - \eta^* b_2)} e^{-it(b_1^+ b_1 + b_2^+ b_2)}, \quad (3)$$

where $H_0 = \hbar\omega_0 a^+ a$.

We suppose that the initial state of the system is $|n>_{cavity} \otimes |0>_{m1} \otimes |0>_{m2}$ where $n > 0$. Employing equ.(3), we obtain the system state dependent on time, i.e.

$$|\psi(t)> = e^{2i(kn)^2(t-\sin t) - |kn\eta|^2} |n>_{cavity} \otimes (|00> + kn\eta|01> + kn\eta|10>)_m, \quad (4)$$

where $|01>_m$ denotes $|0>_{m1} \otimes |1>_{m2}$, others are similar. To acquire the above result, we have employed the condition $gn/\omega \ll 1$, i.e. The condition $gn/\omega \ll 1$ is probably fulfilled experimentally. Actually the experimentally feasible parameters is listed here, $\omega_0 \sim 10^{16} Hz$, $\omega \sim 10^3 Hz$, $L \sim 1m$ and $m \sim 10^{-5} kg$, so $k = g/\omega = \frac{\omega_0}{L}\sqrt{\frac{\hbar}{2m\omega^3}} \sim 10^{-4}$. Even if the photon number in the cavity field is about several hundred, the condition $gn/\omega \ll 1$ is still guaranteed.

### 3. Results and Discussion

In the standard product basis $|1>_m = |1>_m \otimes |1>_m \equiv |11>_m$, $|2>_m = |10>_m$, $|3>_m = |01>_m$, $|4>_m = |00>_m$, the reduced density matrix of the two mirrors is given by

$$\rho = tr_{cavity}(|\psi><\psi|) = e^{-2|kn\eta|^2}\begin{pmatrix} 0 & 0 & 0 & 0 \\ 0 & |kn\eta|^2 & |kn\eta|^2 & kn\eta \\ 0 & |kn\eta|^2 & |kn\eta|^2 & kn\eta \\ 0 & kn\eta^* & kn\eta^* & 1 \end{pmatrix}. \qquad (5)$$

We define matrix $\zeta$ :

$$\zeta \equiv \rho(\sigma_y^1 \otimes \sigma_y^2)\rho^*(\sigma_y^1 \otimes \sigma_y^2), \qquad (6)$$

where $\rho^*$ denotes the complex conjugation of $\rho$ in the standard basis and $\sigma_y$ is Pauli spin matrix (pure imaginary) in the same basis. Wootters concurrence to measure entanglement is given by[18, 19]

$$C(\rho) = \max(0, \sqrt{\lambda_1} - \sqrt{\lambda_2} - \sqrt{\lambda_3} - \sqrt{\lambda_4}), \qquad (7)$$

where $\lambda_1$ is the largest eigenvalue of the matrix $\zeta$. Wootters concurrence varies from C=0 for a disentangled state to C=1 for a maximally entangled state. Substituting equ.(5) into equ.(6), we obtain the only eigenvalue of matrix $\zeta$, $\lambda = 4|kn\eta|^4 e^{-4|kn\eta|^2}$. From equ.(7) Wootters concurrence is given by

$$C(\rho) = \sqrt{\lambda_1} = 2|kn\eta|^2 e^{-2|kn\eta|^2} = 4(kn)^2(1-\cos t)e^{-4(kn)^2(1-\cos t)}. \qquad (8)$$

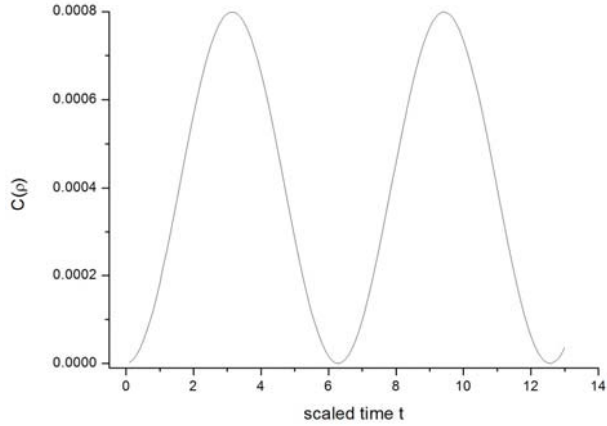

Fig.1 Wootters concurrence $C(\rho)$ vs the scaled time t, with $kn = 0.01$.

Without loss of generality, we show Wootters concurrence vs the scaled time in Fig.1, where $kn = 10^{-2}$. From Fig.1 we obtain three main results: (a) The concurrence does not always equal zero. It means that the two movable mirrors are entangled in the states $|0>, |1>$ via radiation pressure, even the mirrors are typical macroscopic objects. (b) The concurrence is very small, and is about $10^{-4}$ orders of magnitude, from equ.(8) we know that the fact is due to the condition $kn \ll 1$. If the condition $kn \ll 1$ is not fulfilled, the entanglement information may not be extracted. (c) The concurrence dependent on the scaled time is near to a trigonometric function, this fact is easily understood. Because of the condition $kn << 1$, $C(\rho) = 4(kn)^2(1-\cos t)e^{-4(kn)^2(1-\cos t)}$ tends to $4(kn)^2(1-\cos t)$.

## 4. Summary

In conclusion, we have studied the entanglement of two movable mirrors interaction with one single mode cavity field which initial state is $|n>_{cavity} \otimes |0>_{m1} \otimes |0>_{m2}$, worked out Wootters concurrence of the two mirrors without considering thermal noise and classical field driving. We find that the concurrence does not equal zero, so the entanglement of the two movable mirrors does always exist, even the entanglement is very small, about $10^{-4}$. By the way, if the initial state of the cavity field is in a coherent state, there is not entanglement between the two movable mirrors.